\documentclass{aa}
\usepackage{graphicx}
\usepackage{txfonts}
\begin{document} 

\title{Near IR Spectroscopy of Active Galactic Nuclei.\thanks{Based on observations collected at the 
Very Large Telescope (UT1) of the European Southern Observatory, 
Paranal, Chile, ESO NO 63.A-0366}
}               
\author
{C. Boisson\inst{1}
\and S. Coup\'e\inst{1}
\and J. G. Cuby\inst{2}
\and M. Joly\inst{1}
\and M. J. Ward\inst{3}}

\offprints{C. Boisson, \email{catherine.boisson@obspm.fr}}

\institute
    {LUTH, FRE 2462 du CNRS, associ\'e \`a l'Universit\'e Denis Diderot,
    Observatoire de Paris, Section de Meudon\\
 F--92195 Meudon Cedex
\and European Southern Observatory, Santiago, Chile
\and X-ray Astronomy Group, Department of Physics and Astronomy,
      University of Leicester, Leicester LEI 7RH
}

\date{Received  / Accepted  }

\titlerunning{Nuclear and Circumnuclear Near IR Observations of AGN 
}\authorrunning{C. Boisson et al.} 
 
\abstract{ Using the VLT together with the near infrared instrument
ISAAC, we have obtained medium spectral and high spatial resolution
observations of a sample of nearby Seyfert galaxies in the
H-band. This band is particularly suited for stellar population
studies since the stellar component dominates over the AGN
nucleus. The H-band also includes the peak contribution from cool
stars. The AGN spectra are very rich in strong metallic lines which
are sensitive to stellar luminosity class.\\ 
For 4 out of 5 galaxies the central velocity dispersions are found
to be significantly lower than reported in previous
studies. Gradients in the stellar population within the central
regions were searched for, together with evidence for dilution of
the stellar spectral features within the nucleus.\\
\keywords{
Galaxies: stellar content -- Galaxies: active -- Methods: data analysis 
-- Infrared: galaxies}
 }
 
\maketitle\section{Introduction} 
 
An important unsolved issue in understanding the Seyfert phenomenon is 
the relation between the host galaxy and the properties of activity 
within the nucleus. 

 A number of studies carried out at many wavelengths have identified a
connection between the presence of an active nucleus and the presence
of strong star formation within the host galaxy.  These studies found
evidence for young or intermediate age populations in Seyfert~2
galaxies which contribute significantly to the central near-infrared
emission (e.g. Oliva et al., 1999; Heckman et al., 1997;
Gonzalez~Delgado et al., 2001).

The existence of a relationship between stellar population and
activity type was suggested based on results of our study of the stellar
population of a sample of 12 galaxies covering a range of morphology types and
nuclear activities (Boisson et al., 2000). The picture emerging from
our stellar population synthesis is that of an evolutionary sequence
rather than a strict unified scheme for Active Galactic Nuclei
(AGN). Indeed, differences in the stellar populations within the
nuclear regions are found to follow the degree of nuclear
activity. Seyfert 2 galaxies appear to exhibit fossil star formation
activity, i.e. a stellar population younger than found in normal
galaxies, but somewhat older and more metal-rich than found in
starburst galaxies. LINERs have evolved metal-rich stellar populations
with no recent star formation. These findings may suggest intrinsic
differences between galaxies of different degrees of activity. In
particular the Seyfert~2s could be at an earlier stage in the
evolutionary development of low luminosity AGN.

It is therefore of primary importance to define the stellar
populations both within the nucleus and in the surrounding regions of
nearby AGN, with good spatial resolution.

It is also important to cover a wide spectral range in 
order to include a large number of spectral features sensitive to 
various stellar atmosphere parameters.  The spectral variations are 
generally complex and therefore many different lines must to be 
considered in order to avoid ambiguities in the interpretation.  As an 
example, the near-IR CaII triplet and the IR CO and Si indices can be 
strong {\em not only} in supergiant stars {\em but also} in metal rich 
giants; colour gradients can be caused either by changing stellar 
populations or by dust effects. For these and other reasons stellar 
population synthesis is better constrained by using observations over 
a broad wavelength range.

The near-infrared range is a very useful 
domain to study the stellar content of active galaxies. In particular 
the H-band is well suited as the non-stellar contribution (mainly 
reprocessed nuclear emission by dust) is much less than in the K 
window. For example, it has been shown by Origlia et al. (1993) that 
in the Seyfert~2 nucleus of NGC1068 the non-stellar contribution is 
only 30$\%$ within the H-band whereas it is about 80$\%$ in the 
K-band. The H-band is also the region in which emission from cool 
stars peak, and the extinction by dust is lower than for the shorter 
wavelength J-band. 

Moreover, restricting H-band observations to the region
1.57--1.64~$\mu$m, which is free of strong AGN emission lines, allows
us to sample the stellar content of the very nucleus, unlike the
situation in the optical for type~1 objects, where strong broad
emission lines seriously mask the stellar features.  Dallier et
al. (1996) have pointed out the presence, in the H-band, of very good
luminosity discriminators for stars later than K0. The advantage of
these discriminators is that the relevant lines lie close together
in wavelength, so that they can be observed at the same time
and furthermore their ratio is not affected by dilution from
dust emission.

In this paper we present medium resolution ($R\sim3300$) H-band
spectroscopy for five Seyfert galaxies of type 1 and type 2. The
spectra of several stars obtained under the same conditions are also
presented; these serve to extend the stellar library, for the same
wavelength range and at a similar spectral resolution, as previously
published by Dallier et al. (1996), for the purpose of stellar
population synthesis. The observations and data reduction are
described in Sect.~2, results are reported in Sect.~3 and a short
discussion is given Sect.~4. The star spectra are presented in
Appendix~A.
 
\section{Observations and Data Reduction} 
 
We obtained, in service mode, long slit $H$-band spectroscopy of the
central regions of a sample of AGN using the ISAAC specrograph mounted
on the VLT/ANTU telescope (Cuby et al., 2000; Moorwood et al.,
1999). The spectral region, 1.57--1.64~$\mu$m, covered using one
grating position includes many metallic stellar features. During the
observations the seeing varied between 0.6 and 1 arcsec.  Using a slit
width of 1 arcsec we obtained a spectral resolution of 4.5\AA\ FWHM
with a spectral sampling of 0.79\AA. The spatial sampling is 0.147
arcsec/pixel.

In Table~1 we list the observed galaxies, their morphological and
Seyfert type, redshift, distance in Mpc (assuming H$_o$=75km/s/Mpc and
q$_o$=0.0), colour excess due to Galactic interstellar reddening and
the velocity dispersion within the central region together with its
publication reference. The values of E(B-V) have been evaluated using
the Galactic hydrogen column density derived from the 21cm survey of
Schlegel et al. (1998).

 Each galaxy was observed for a total of 3hrs (DIT=600sec. $\times$
2 (for position $A$ and $B$) $\times$ 3 AB cycles $\times$ 3 =
3hrs). We observed using the nodding mode: the objects in exposures
``$A$'' and``$B$'' were centered 120 arcsec apart, the observed
infrared extent of the sample galaxies being less than that (the sky
emission dominated beyond an extent of 10 arcsec). This set-up
allowed us to use the differential comparison ($A-B$ and $B-A$) to
subtract the dark, bias and sky contribution from all exposures.  We
also observed a set of early type stars to use in the removal of
telluric features, and a set of cool stars for the flux calibration.

\begin{table*}
\caption[]{Observed galaxies.}
\bigskip
\begin{tabular}{|l|c|c|l|c|c|l|c|c|}
\hline
\noalign{\smallskip}
galaxy &morphology& Seyfert &z& D   & E(B-V) & Epoch &$\sigma_{lit}$& Ref.\\

       &          & type    & & Mpc &        &  &(km s$^{-1}$) & \\
\noalign{\smallskip}
\hline
\noalign{\smallskip}

NGC 2992     & Sa  & 2 & 0.0077  & 31 & 0.06 &05/04/1999   &158$\pm$13 & NW \\
~            &     &   &         &    &      &             &188-218    & OOMM\\
NGC 3185     & SBa & 2 & 0.0041  & 16 & 0.03 &05/08/1999   &61$\pm$20  & NW \\
NGC 3783     & SBa & 1 & 0.0098  & 40 & 0.12 &05/02/1999   &138-165    & OOKM\\
NGC 6221     & SBbc& 2 & 0.0047  & 19 & 0.17 &05/01/1999   &120-140    & VB\\
MCG-06-30-015& E   & 1 & 0.0077  & 31 & 0.06 &05/05/1999   &153-165    & OOMM\\

\noalign{\smallskip}
\hline
\end{tabular}
\smallskip

Ref.: NW: Nelson \& Whittle, 1995; OOMM: Oliva et al., 1999; OOKM: Oliva 
et al., 1995; VB: Vega-Beltr\'an et al., 1998. 
\end{table*}

\begin{table*}
\caption[]{Observed stars.}
\bigskip
\begin{tabular}{|l|c|c|c|}
\hline
\noalign{\smallskip}
star & type & [Fe/H] & Ref. \\

\noalign{\smallskip}
\hline
\noalign{\smallskip}

HD  11507  & M1V  &  -   &  \\
HD  36395  & M1V  & 0.60 & CS \\
HD  39715  & K3V  & 0.33 & BG \\
HD  48501  & F2V  & 0.01 & CS \\
HD  93250  & O6   &  -   &  \\
HD 106116  & G4V  & 0.15 & FMS \\
HD 112164  & G1V  & 0.24 & CS \\
HD 121790  & B2V  &  -   &  \\
HD 131977  & K4V  & 0.01 & CS \\
HD 139717  & F8Ib & 0.18 & SKC \\
HD 159217  & A0V  &  -   &  \\

\noalign{\smallskip}
\hline
\end{tabular}
\smallskip

Ref.: BG: Barbuy \& Grenon, 1990; CS: Cayrel de Strobel et al., 1997;
FMS: Favata et al., 1997; SKC: Soubiran et al., 1998.
\end{table*}

In Table~2 we list the 
stars observed, their spectral type and luminosity class and, where 
known, the metallicity together with the published reference.  A dash symbol 
for the metallicity indicates that no [Fe/H] information is 
available. The spectra are presented in Annex A.

 There follows a brief description of the reduction and analysis
procedures applied to our data using {\it MIDAS, IRAF} and also some
additional routines from an early version of the {\it Eclipse} reduction
package (Devillard, 1997).

The data were first corrected for the ``electrical ghost'' features
generated by the detector, using the is-ghost routine provided in the
{\it Eclipse} package. The data were then flat-fielded using a master
flat field image (this pixel to pixel correction is particularly
critical for our data as the frames show features which disappeared
after the major "repair" of ISAAC in February 2000). Using the {\it
IRAF} package we then corrected for distortions along and
perpendicular to the slit, using the nuclei of the galaxies together
with the calibration stars observed, as no star-trace exposures were
provided by ESO at that time. Individual exposures are then combined
using the {\it spjitter} routine of the {\it Eclipse} package.

Throughout the spectral region of our observations,
there are many bright OH lines that
can be used to perform a wavelength calibration of the exposures. This
calibration was carried out using the MIDAS package as it was found to
be non-linear. Precision to about 1/7th of a pixel was achieved.

The 1-D spectra were then extracted from the 2-D frames and corrected
for telluric absorption using the early type stellar templates,
taking into account the differential airmass. An example of the
telluric spectrum is shown in Fig.~1. The overall shape of stellar
spectra of the early type and of cool stars we observed were compared
to the model spectra taken from Hauschildt et al. (1999a,b) in order
to derive relative flux calibrations. The final stellar spectra were
found to be in agreement within a few percent, over the full
wavelength range, with those from our stellar library
(Dallier et al, 1996). Photometric calibration was achieved by using
aperture magnitudes taken from the literature (Kotilainen et al., 1992;
Glass \& Moorwood, 1985) and by appropriate scaling of the extracted
spectra for the corresponding aperture. No photometric data could be found
for NGC~3185, and so only relative fluxes are displayed.

\begin{figure}
\resizebox{\hsize}{!}{\includegraphics[angle=-90]{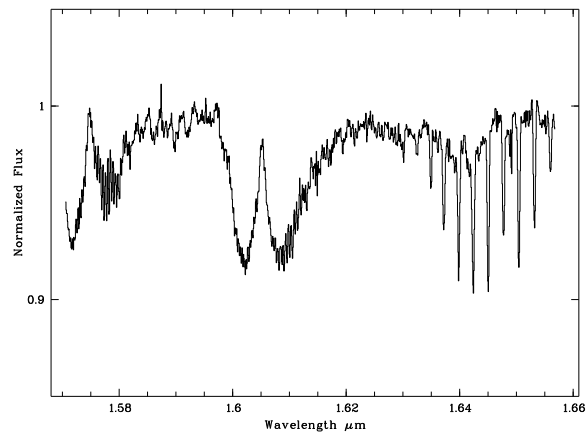}}
\caption{Telluric absorption spectrum.}
\end{figure}

Logarithmic flux profiles across the galaxy 2D-spectra (Fig.~2) were
used to determine where to extract the 1D spectra of the nucleus and
the regions around it. The optimum number of rows to be summed depends on
the seeing. When the spectra of two symmetric regions lying on either
side of the nucleus were found to be very similar, indicating that the
stellar population of a galaxy is homogeneous, we averaged them
together to obtain a mean spectrum.  Figures~3 to 7 show the spectra
extracted from the five galaxy frames: the nucleus and the surrounding
regions. All of the spectra displayed are corrected to the rest wavelengths.

\begin{figure*}
\includegraphics[width=12cm,angle=-90]{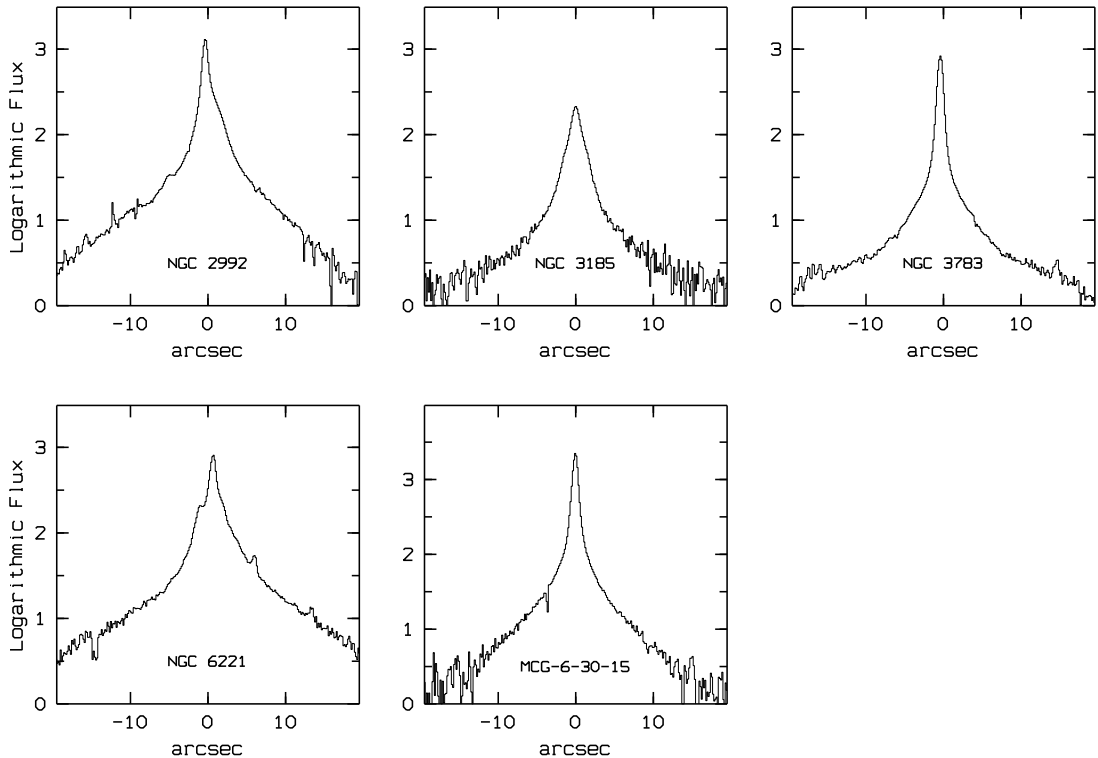}
\caption{Logarithmic profiles across the galaxy 2D-spectra.}
\end{figure*}

\begin{figure*}
\resizebox{\hsize}{!}{\includegraphics[angle=-90]{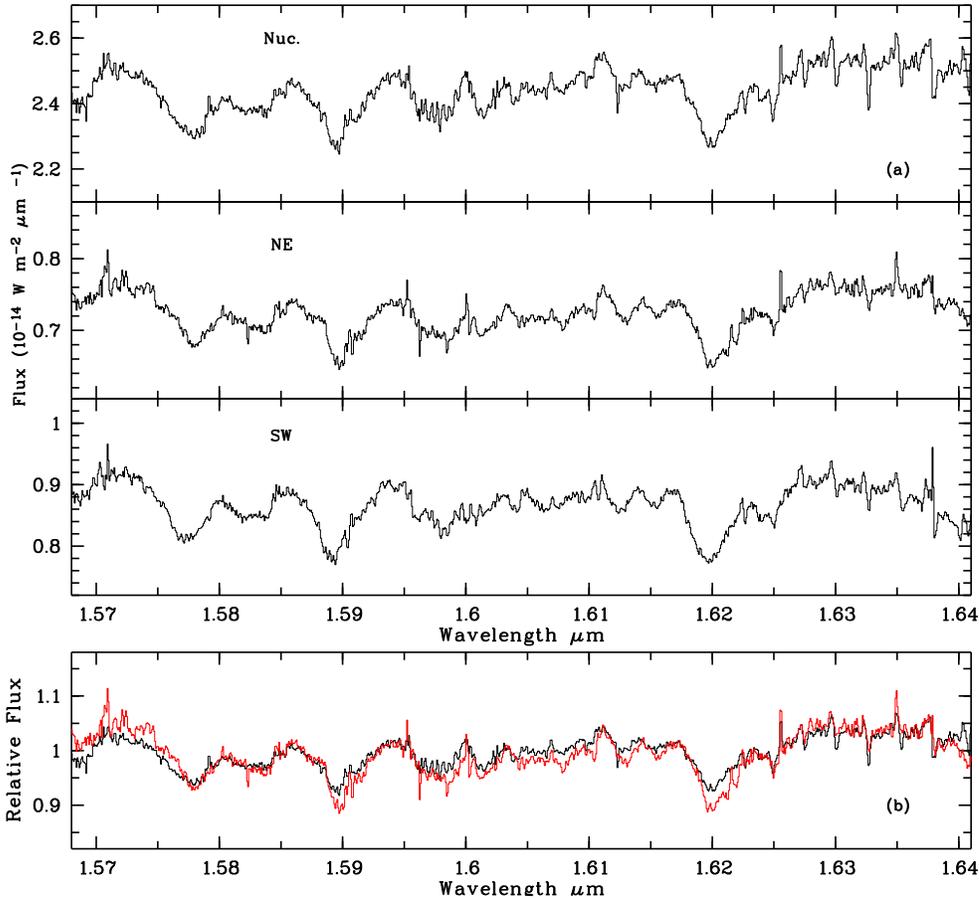}}
\caption{NGC~2992: (a) Spectra of the nucleus, and of two off-nucleus
 regions located at 225 pc from the centre.  (b) Spectrum of the NE
 region (grey line) shifted by 15 km s$^{-1}$ and scaled in flux, is
 superimposed on the nuclear spectrum (black line). Wavelengths are in
 the rest frame of the galaxy.}
\end{figure*}

\begin{figure*}
\centering
\resizebox{\hsize}{!}{\includegraphics[angle=-90]{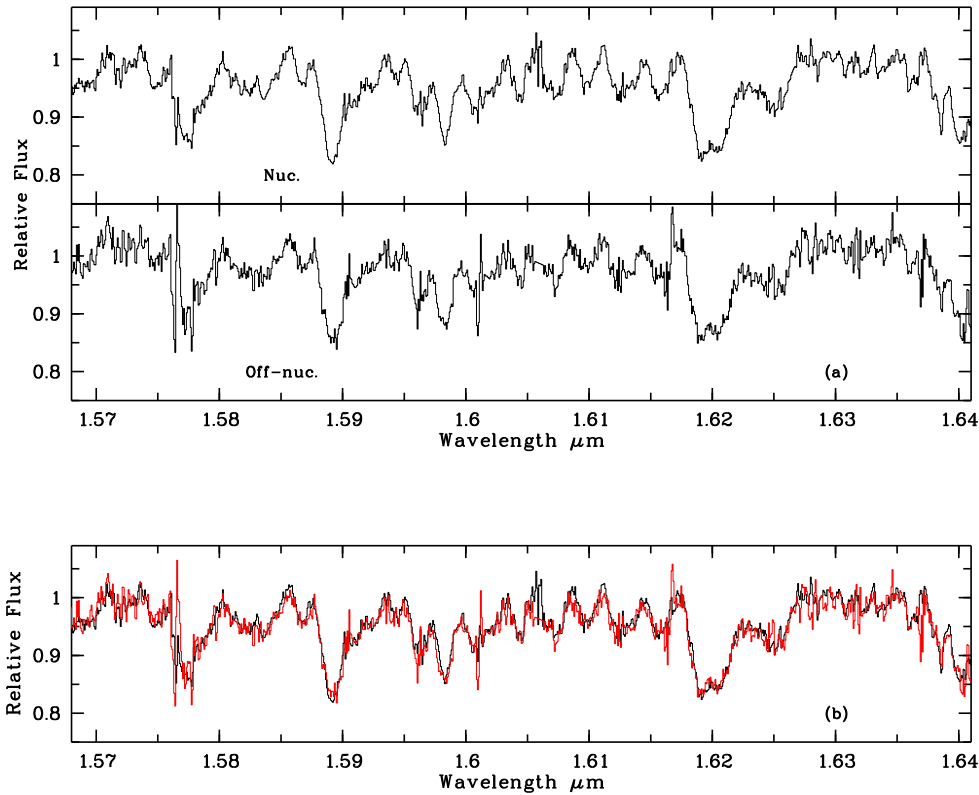}}
\caption{NGC~3185: Top - Nuclear spectrum, Bottom - Off-nucleus
spectrum of a ring region centred at 120 pc from the
nucleus. Wavelengths are in the rest frame of the galaxy. }
\end{figure*}

\begin{figure*}
\includegraphics[width=12cm,angle=-90]{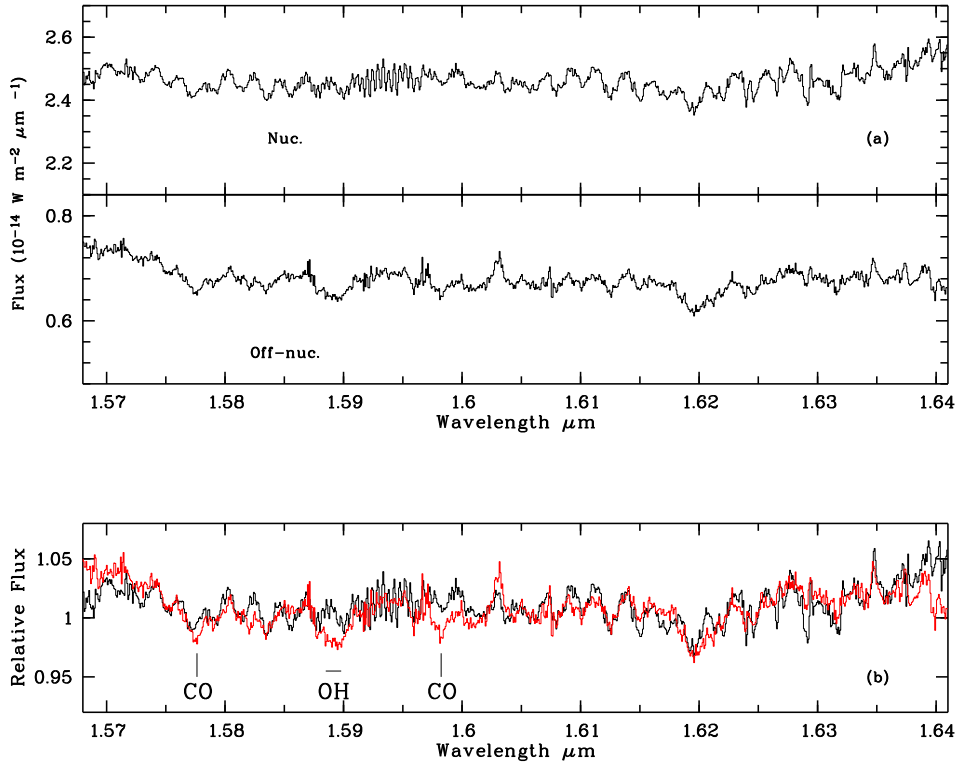}
\caption{NGC~3783: (a) Spectra of the nucleus and a ring region
centred at 280 pc from the nucleus.  (b) A composite spectrum of 50\%
stellar ring component and 50\% hot dust emission (grey line)
superimposed on the nuclear spectrum (black line). Wavelengths are in
the rest frame of the galaxy.}
\end{figure*}

\begin{figure*}
\includegraphics[width=17cm,angle=-90]{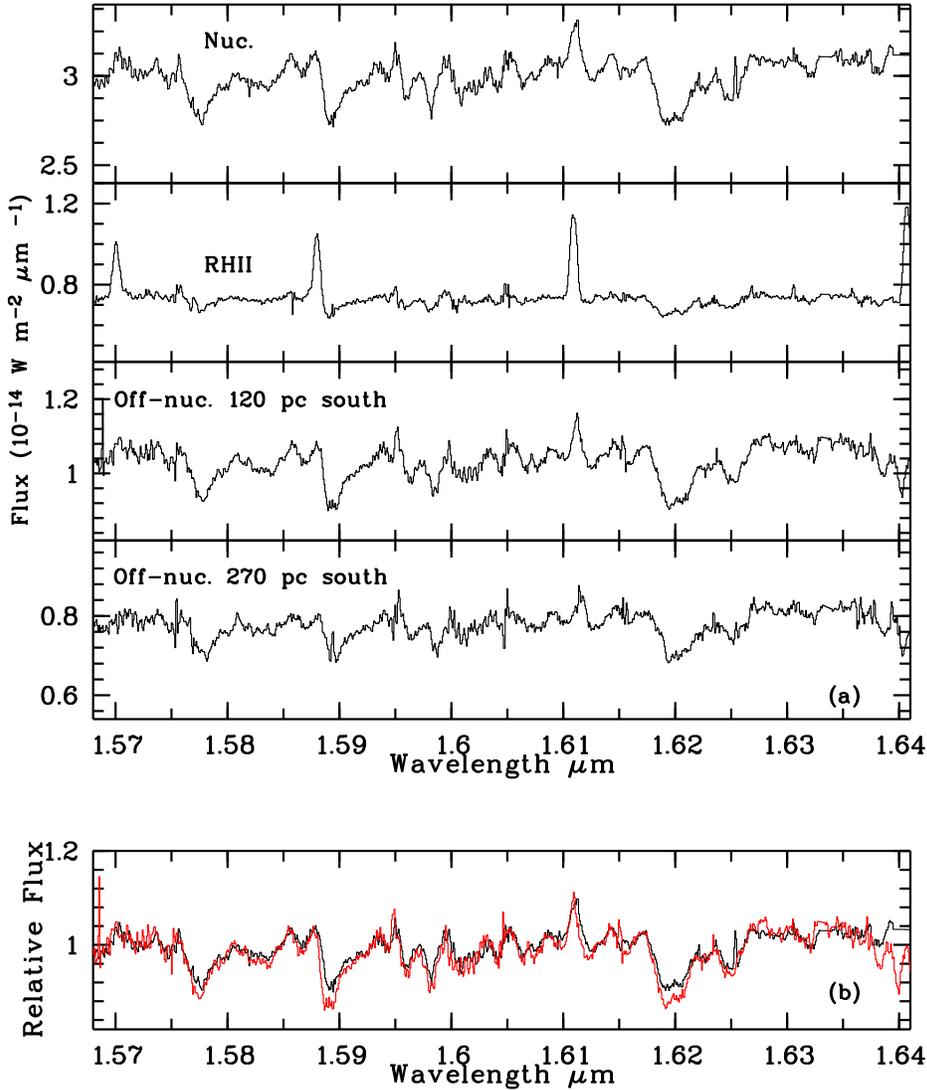}
\caption{NGC~6221: From top to bottom - Nucleus, HII region 
located 150 pc north of the centre, two off-nucleus regions south of the
centre. Wavelengths are in the rest frame of the galaxy. }
\end{figure*}

\begin{figure*}
\includegraphics[width=12cm,angle=-90]{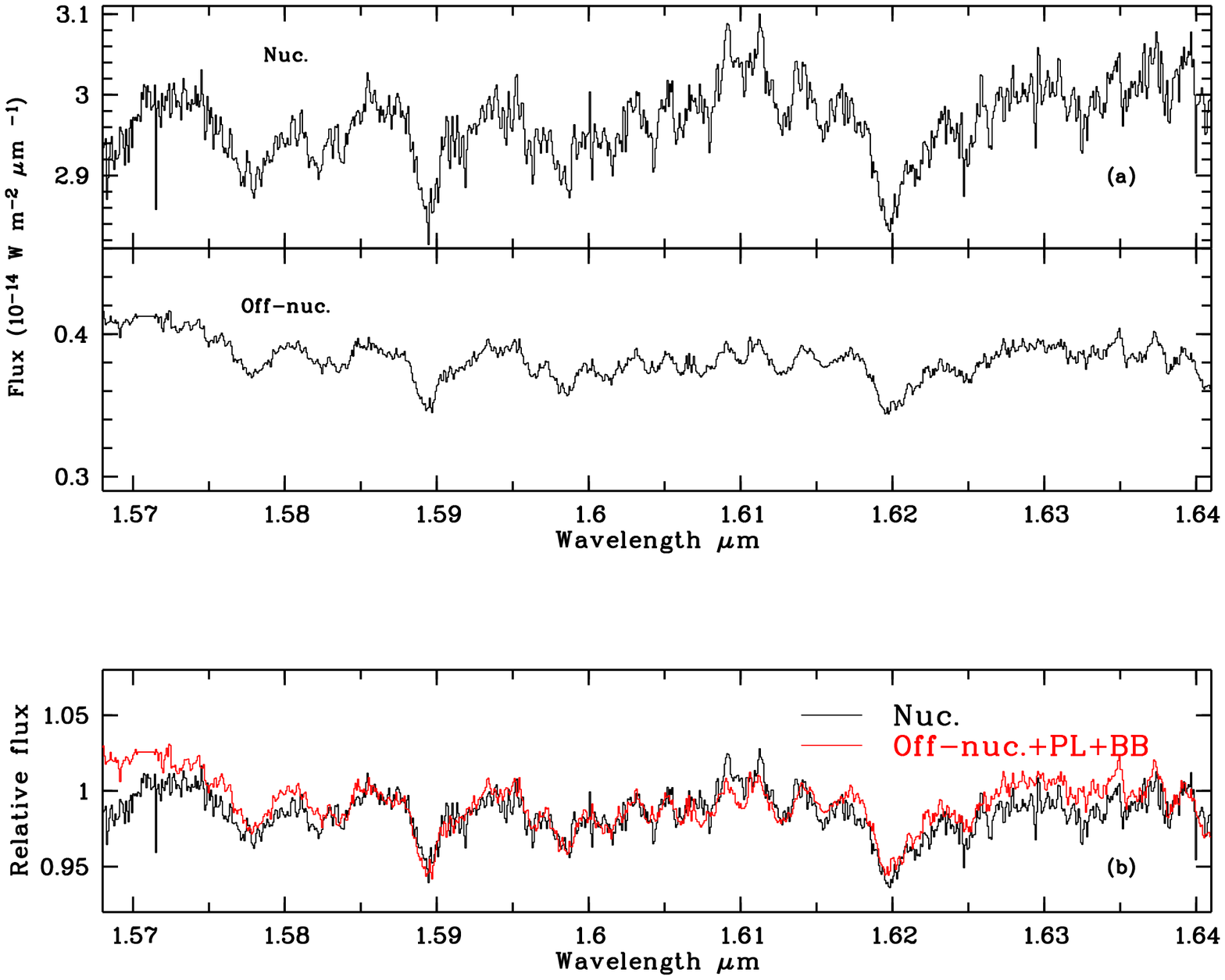}
\caption{MCG-06-30-15: (a) Spectra of the nucleus and an off-nucleus
ring region located at 220 pc from the centre.  (b) A composite
spectrum of 50\% stellar ring component, 30\% warm dust and 20\%
power-law continuum emission (grey line) superimposed on the nuclear
spectrum derreddened by E(B-V)=0.05 (black line). Wavelengths are in
the rest frame of the galaxy.}
\end{figure*}

A detailed identification of the absorption features
observed in the stars and the galaxies was carried out using the
solar spectrum published by Livingston \& Wallace (1991). The observed
wavelengths and identifications are listed in Table~3. Numerous
lines of neutral species are present, in addition to broad features
due to the CO and OH absorption bands, as can be clearly seen in Figures 3 to
7.

\begin{table*}
\caption[]{Line identification Table}
\bigskip
\begin{tabular}{|c|l|c|l|}
\hline
\noalign{\smallskip}
$\lambda_{\mu}$ & Identification&$\lambda_{\mu}$ & Identification \\

\noalign{\smallskip}
\hline
\noalign{\smallskip}

1.5701  & Br(4-15) &                            1.6072  & FeI 1.6072 + OH 1.6070-1.6074 \\
1.5726  & FeI 1.5724 + OH 1.5727-1.5730 &       1.6095  & SiI 1.6095 \\                   
1.5743  & MgI 1.5741 &                          1.6102  & FeI 1.6102 \\                   
1.5751  & MgI 1.5749 &                          1.6112  & Br(4-13) \\                     
1.5758  & OH(3,3)P(7) 1.5756 &                  1.6116  & FeI 1.6117 \\                   
1.5768  & MgI 1.5766 + FeI 1.5770 &             1.6126  & FeI 1.6126 + OH 1.6124 \\       
1.5776  & CO(4-1) 1.5775-1.5784 + OH 1.5778 &   1.6136  & NiI 1.6136 + CaI 1.6137 \\      
1.5791  & FeI 1.5789 &                          1.6153  & CaI 1.6150-1.6158 + FeI 1.6153 \\
1.5800  & FeI 1.5799 &                  1.6163  & FeI 1.6165 + SiI 1.6164 \\      
1.5812  & FeI 1.5810 &                  1.6180  & FeI 1.6180 \\                   
1.5820  & FeI 1.5820 &                  1.6186  & CO(6,3) 1.6186 \\               
1.5837  & SiI 1.5834 + FeI 1.5834-1.5840 &      1.6191  & OH 1.6190-1.6192 \\             
1.5856  & FeI 1.5854 &                          1.6197  & CaI 1.6198 + FeI 1.6198 \\      
1.5866  & FeI 1.5868 + OH 1.5864 &              1.6205  & FeI 1.6208 + OH 1.6204-1.6207 \\
1.5871  & FeI 1.5869 &                          1.6214  & SiI 1.6216 + FeI 1.6214 \\      
1.5881  & FeI 1.5878 + MgI 1.5880 &             1.6224  & $^{13}$CO +OH 1.6224-1.6226 \\  
1.5884  & Br(4-14) &                            1.6231  & FeI 1.6232 + OH 1.6216-1.6231 \\
1.5890  & SiI 1.5888 + OH 1.5885-1.5888 &       1.6241  & SiI 1.6242  \\                  
1.5895  & FeI 1.5895 + OH 1.5892-1.5898 &       1.6252  & OH 1.6241-1.6260 \\             
1.5900  & FeI 1.5898 &                          1.6270  & OH 1.6266-1.6276 \\             
1.5907  & FeI 1.5905+1.5906 &                   1.6283  & FeI 1.6285 \\                   
1.5913  & OH 1.5910-1.5913 &                    1.6291  & FeI 1.6292 \\                   
1.5922  & FeI 1.5921 &                          1.6299  & OH 1.6298-1.6300 \\             
1.5942  & FeI 1.5942 &                          1.6317  & FeI 1.6316 + OH 1.6312-1.6316 \\
1.5956  & FeI 1.5954 + MgI 1.5955 &             1.6324  & FeI 1.6325 \\                   
1.5961  & SiI 1.5960 &                          1.6332  & FeI 1.6332 \\                   
1.5967  & FeI 1.5965+1.5968 + OH 1.5968 &       1.6347  & OH 1.6346-1.6348 \\             
1.5982  & CO(5,2) 1.5978 + FeI 1.5981 &         1.6354  & OH 1.6352-1.6355 \\             
1.5999  & FeI 1.5998 &                          1.6365  & NiI 1.6363 + OH 1.6365 \\       
1.6008  & FeI 1.6008 &                          1.6381  & SiI 1.6381 + FeI 1.6382 \\      
1.6017  & FeI 1.6017 &                          1.6396  & FeI 1.6396 \\                   
1.6038  & OH(3,1)P(9) 1.6038 &                  1.6399  & CO(7,4) 1.6396 \\               
1.6041  & FeI 1.6041 &                          1.6406  & FeI 1.6405 + Br(4-12) \\        
1.6061  & SiI 1.6060 + OH 1.6053-1.6065 &       1.6445  & FeI 1.6445 + OH 1.6443-1.6450 \\
\noalign{\smallskip}
\hline
\end{tabular}
\end{table*}

\section{ Results}

High spectral resolution observations such as this can be used to measure
velocity dispersions at different locations within the central regions of
the galaxies.

We have employed two different methods to determine stellar velocity
dispersions. The first method is similar to that described in Tonry \&
Davis (1979) i.e. cross-correlation of the galaxy spectrum with
template stars is fitted using the autocorrelation function of the
star convolved with a Gaussian, with $\sigma$ and peak velocity as
free parameters.  The second method consists of fitting the galaxy
spectrum with a template stellar spectrum convolved with a Gaussian.
In this case the fit parameters are the $\sigma$ of the Gaussian
profile and a multiplicative factor to fit the depth of the stellar
features. In both cases fitting is performed over the whole wavelength
range, using different K and M stellar templates in order to minimize
uncertainties introduced by a possible mismatch between the stellar
template and the observed galaxy's stellar population. The use of
different K and M stellar templates also allows to account for the
errors due to seeing variations between two observations.

The two fitting methods give consistent results. The velocity
dispersions and final one sigma errors are displayed in Table~4. These
velocity dispersions are quite low. Also, the stellar velocity
dispersions are roughly constant in the central regions sampled
here. We note that stellar lines are significantly diluted within the
nucleus of NGC~3783, and so no velocity dispersion could be derived.

Except for NGC~6221, these low values are in contrast with the bulge
velocity dispersions previously published (see Table~1). A similar situation
is found for Circinus, with a velocity dispersion of 80 km/s
in the central 80 pc observed using 3D (Maiolino et al., 1998) while a
value of 168 km/s over the same region was deduced from IRSPEC
observations (Oliva et al., 1995). We suggest that these differences
indicate an improvement in the measurement of the velocity
dispersions, and is due to our higher spectral resolution as well as
to the large number of stellar lines sampled in the fitting
procedure. Part of the difference between the values measured here using 
a small aperture compared to previous ones measured in larger apertures
may be due to the presence of a rotational component which contributes to 
the breadth of the features. However recent kinematical profiles of the
hosts of some AGN (Emsellem et al., 2001) obtained with ISAAC show that the
dispersion remains nearly constant inside the nuclear disk, with low
values. For example, in the case of NGC~1808 the dispersion profiles
remain in the range 80--120 km s$^{-1}$ over the central 20 arcsec.,
while Oliva et al. (1995) measured 154 km s$^{-1}$ in a $\sim$ 5
arcsec aperture.

The equivalent widths (EW) of strong lines have been commonly used to
estimate the relative contribution of AGN and stellar emission in the
nucleus of Seyfert galaxies. Following the work of Oliva et al. (1999) in
Table~4 we give the EWs for the two stellar features at 1.59$\mu$m and
1.62$\mu$m, measured in the nuclear and the off-nuclear spectra
of the 5 AGN. These EWs have been measured for the same wavelength
domain as discribed in Origlia et al. (1993).

\begin{table*}
\caption[]{Velocity dispersion.}
\bigskip
\begin{tabular}{|l|l|r|l|c|c|}
\hline
\noalign{\smallskip}
galaxy & &R$^*$  & $\sigma$ & EW(1.59$\mu$m) & EW(1.62$\mu$m)\\

       & &  pc   & (km s$^{-1}$) &\AA &\AA \\
\noalign{\smallskip}
\hline
\noalign{\smallskip}

NGC 2992     & nuc. & 110 &95$\pm$18 &2.7&3.2 \\
~            & SW   & 225 &79$\pm$11 &3.6&4.5 \\
~            & NE   & 225 &88$\pm$15 &3.8&4.6 \\
NGC 3185     & nuc. & 60  &75$\pm$13 &4.2&5.8 \\
~            & ring & 120 &81$\pm$15 &3.9&5.4 \\
NGC 3783     & nuc. & 140 &      -   &0.9&1.6 \\
~            & ring & 280 &98$\pm$19 &1.7&3.6 \\
NGC 6221     & nuc. & 70  &63$\pm$9  &2.3&3.9 \\
~            & r1   & 120 &68$\pm$9  &3.8&5.5 \\  
~            & r2   & 270 &71$\pm$10 &$>$3.3&5.1\\  
MCG-06-30-015& nuc. & 100 &69$\pm$14 &1.1&1.8 \\
~            & ring & 200 &83$\pm$12 &2.8&3.7\\ 

\noalign{\smallskip}
\hline
\end{tabular}
\smallskip

* radius for the nuclear region; distance to the center for the
off-nuclear region
\end{table*}

\subsection{NGC 2992} 
 
The Seyfert 1.9 galaxy NGC~2992 is a Sa galaxy seen almost edge-on and 
is interacting with NGC~2993. A prominent dust lane extending along 
the major axis crosses the nucleus of the galaxy (Ward et al., 
1980). This galaxy exhibits a biconical galactic-scale outflow, which 
emerges almost perpendicularly from the plane of the galaxy (Veilleux 
et al., 2001 and references therein). 

From a comparison with Bica's (1988) templates, Storchi-Bergmann et
al. (1990) found an old stellar population of solar metallicity coupled
with a small contribution ($\leq$ 5\%) of recent star formation in the
inner 5 arcsec. Oliva et al. (1999) deduced from comparison of the
features at 1.59$\mu$m and 1.62$\mu$m, that the flux from inner 4 arcsec is
largely dominated by stellar light with no significant contribution from a
starburst. In a more detailed IR adaptive optics study of NGC~2992
Chapman et al. (2000) found that the radial distribution of
the CO index indicated a strong population gradient within the core,
with the stellar population at the very centre being older than that in the
surrounding regions.

The flux profile along PA=55$\degr$ (Fig.~2) shows a distinct nucleus
together with an asymmetric galaxy component. In the inner 5 arcsec,
the NE region may be partly absorbed by the dust lane.

The flux extracted from regions 1.5 arcsec (1'' $\sim$ 150pc) from the
nucleus (Fig.~3a) have similar stellar spectra but with
different kinematics. The emission line gas is known to exhibit two
distinct kinematic components, one due to galactic rotation and another
resulting from outflow (e.g. M\'arquez et al., 1998). The stellar features from
the SW region are redshifted by 50$\pm$5~km~s$^{-1}$, and those from the
NE region are blueshifted by 15$\pm$4~km~s$^{-1}$ with respect to the
nucleus, in agreement with the gaseous component of rotation. 

The similarity of the spectra from both regions implies an homogeneous
stellar population within the bulge of this galaxy. The nucleus of
this galaxy is very variable, possibly related to a supernova event
(Glass, 1997), and the hot dust component appears to be in the process
of increasing at the time of our observations (Gilli et al.,
2000). The fraction of stellar light in the nucleus, for the
hypothesis of dilution by AGN and/or hot dust emission, is deduced
from the comparison of the nuclear EW to that of the surrounding
regions. From the EWs in Table~4 we infer up to 30\% dilution.  But
such an analysis can be misleading. Indeed, comparison of the nuclear
and off-nucleus spectra {\it over the full wavelength range} show that
only the OH lines around 1.59$\mu$m and 1.62$\mu$m are weaker on the
nucleus, while no gradient in line strength is found for the other
elements (Fig.~3b). This favours a stellar population gradient rather
than dilution by a non-stellar component. Although hot dust
emission is not inferred to contribute to the H-band, it is still
possible that it may contribute to the K-band emission as found by
Gilli et al.

\subsection{NGC 3185} 
 
NGC~3185 is a so-called transition object (having both Seyfert and HII
region characteristics) with a host galaxy of SBa type located
in a compact group including NGC~3190 (Gon\c{c}alves et al., 1999).
To our knowledge, there are no published results on the stellar
content of the inner bulge of this galaxy.

The radial flux profile along PA=0$\degr$ of this galaxy is smooth
without a strong nucleus (Fig.~2). In Figure~4a we show the spectra of
the nucleus and that of an extended region obtained by averaging
together similar data extracted from both sides of the nucleus,
1.5~arcsec away (1''$\sim$ 80pc). 

No evidence is found either for a stellar population gradient or dilution
by dust (Figure~4b and Table~4).

\subsection{NGC 3783} 
 
NGC~3783 is a nearly face-on SBa galaxy with a very bright Seyfert~1
nucleus. Winge et al. (1990) found that the stellar population of the
bulge is of solar metallicity and mostly old, but with some contribution
from intermediate age and moderately young stars. 

A strongly peaked nucleus on a faint symmetrical bulge is apparent in
Figure~2. The bulge spectrum (1.5 arsec from the nucleus;
1''$\sim$190pc) exhibits strong stellar absorption lines and the stellar
population appears significantly diluted in the nucleus (see Fig.~5a
and Table~4).

Within a Seyfert~1 nucleus we might expect that the AGN continuum
would dominate over the contribution from the stellar population, and
therefore that the stellar features would be heavily diluted by the
typical power-law or/and hot dust emission from the AGN. From comparison
of the features at 1.59$\mu$m and 1.62$\mu$m, Oliva et al. (1999)
found 55\% stellar contribution to the 4 arcsec nuclear H band
flux. Adding the CO~2.29$\mu$m line information they parametrize the
non-stellar flux in terms of a power-law with $\alpha$=3.6
(F$_{\nu}$=${\nu}^{-\alpha}$) plus a T=930K dust emission. 

Another way of testing the dilution hypothesis is to compare the
continuum shape of the nuclear and off-nuclear spectra. We approximate the
dust emission, within the 1.5'', using blackbody spectra with
temperatures ranging from 600K up to the sublimation temperature of
around 1500K. A reasonable fit to the nuclear spectrum is obtained
with a 50\% stellar contribution diluted by 1500K dust emission (see
Fig.~5b). However a equally good fit can be obtained using a larger
dilution due to both colder dust (600K), plus a typical AGN power-law
contribution ($\alpha$=1); which is expected as dust and power-law
dilutions have opposite wavelength dependence.

A stellar population gradient is probably also marginally present as
the OH and CO lines appear weaker in the nucleus than in the
surrounding regions.

\subsection{NGC 6221} 
 
NGC~6221 is located within a small galaxy group. The host galaxy is a
barred Sbc with a weak Seyfert nucleus. The radio morphology and
spectral index of the nucleus and bar are indicative of the presence
of supernovae remnants (Forbes \& Norris, 1998). This galaxy is known
to exhibit clear signs of intense circumnuclear star formation. The
optical spectrum of the central 6 arcsec shown in Phillips (1979),
exhibits signatures of emission originating from HII regions together
with a weak Seyfert nucleus, whereas the underlying continuum is
apparently dominated by late type stars. Cid~Fernandes et al., (1998)
confirms the presence of a nuclear starburst within the central 5
arcsec, with the circumnuclear regions being somewhat older. Levenson
et al. (2001) shows that even in the central 1.5x1.5 arcsec, the AGN
nucleus is weak relative to surrounding starbursts.

NGC~6221 is the most extended galaxy in our sample, as can be seen
from its logarithmic flux profile along PA=30$\degr$. The conspicuous
peak at 1.7~arcsec north from the nucleus (Fig.~2; 1''$\sim$ 90pc) is
clearly an HII region.

In addition to the HII region mentioned above, three 1-D spectra were
extracted: the nucleus and two regions at 1.3 and 3~arcsec south
respectively from the centre (Fig.~6).  The stellar features from the
regions are blueshifted by 34$\pm$3~km~s$^{-1}$ and
68$\pm$4~km~s$^{-1}$ with respect to the nucleus, in agreement with
the kinematics of the stellar component as published by Vega-Beltr\'an
et al. (1998). 

From EWs in Table~4 we infer up to 30\% of dilution.  As for
NGC~2992, such an analysis can be misleading: comparison of the nuclear
and off-nucleus spectra {\it over the full wavelength range} show that
only the lines around 1.59$\mu$m and 1.62$\mu$m are weaker at the
nucleus while all other lines have similar depths (Fig.~6b).

The strong similarity of both off-nuclear spectra suggests an
homogeneous stellar population in the bulge. However the stellar
component in the very nucleus could be somewhat different. The
Brackett (4-13) line, a signature of HII regions, is also conspicuous
in all three spectra.
 
\subsection{MCG-6-30-15} 
 
The Seyfert~1 MCG-6-30-15 is a very elongated lenticular 
galaxy. Colour maps show a small dust lane south of the nucleus 
roughly parallel to the major axis of the galaxy. The central region 
of the galaxy is much redder than the outer regions (Ferruit et al., 
2000). 

From a stellar population study based on ultraviolet spectra, 
Bonatto et al. (2000) claim that an old bulge stellar population is 
the dominant contributor, with indications of a series of previous 
bursts of star formation.

The logarithmic flux profile shows a distinct nucleus on a weak,
extended galaxy component (Fig.~2). In Figure~7a we show spectra of
the nucleus (1.5~arcsec wide; 1''$\sim$ 150pc) and that of an extended
region formed by averaging together similar data extracted from both
sides of the nucleus, 1.5~arcsec away. The nuclear and circumnuclear
regions have similar spectral features although these are weaker in
the nucleus. The nuclear spectrum is also redder.

Since the nucleus is a Seyfert~1, it is natural to interpret these
differences in terms of dilution by a hot dust component and/or a
power-law AGN type continuum. From comparison of the features at
1.59$\mu$m and 1.62$\mu$m, Oliva et al. (1999) found 40\% stellar
contribution to the 4 arcsec nuclear H band flux. Adding the
CO~2.29$\mu$m line information they parametrize the non-stellar flux
in terms of a power-law with $\alpha$=2.3
(F$_{\nu}$=${\nu}^{-\alpha}$) plus a T=1100K dust emission. Within the
1.5 central arcsec we infer a similar dilution (Table~4).

A strong intrinsic reddening (E(B-V)=0.61 to 1.09) in the very nucleus
has been previously deduced in multi-wavelength study by Reynolds et
al. (1997). Comparing the continuum shape of the full nuclear and
off-nuclear spectra, if the lower end of reddening value range is
assumed, then dilution by both a black-body and a power-law continuum
are needed. However, if a higher value of the reddening is assumed,
then dilution by only the power law is necessary.  Figure~7b shows the
comparison between the nuclear spectrum dereddened by E(B-V)=0.5 and
the circumnuclear spectrum diluted by 20\% power-law continuum and
30\% warm dust.
 
\section{Discussion} 
  
We have presented and discussed H-band spectra at a 1.5~arcsec spatial
resolution of a sample of three Seyfert~2 and two Seyfert~1
galaxies. It appears that the nuclear spectra are dominated by stars
in the Seyfert~2s, while evidence for dilution of their stellar
components by hot dust and/or power-law AGN is found in the cases of
the Seyfert~1s. 

A classical tool used to constrain the luminosity class of the dominant
stars in an integrated population is its mass-to-light ratio. Under
the assumption that the nuclear mass and luminosity of the galaxy
share the same radial dependance (Devereux et al., 1987) and that the
measured velocity dispersion does not vary with the size of the
aperture used, the IR mass-to-light ratio is simply given as a
function of the velocity dispersion, the projected spectrometer
aperture and the 1.65$\mu$m stellar fluxes (see Oliva et al. (1999),
equation 3). M/L ratios computed in this way are very dependant on the
stellar density and not only on the stellar luminosity, a factor that
cannot be ignored especially in these centrally peaked objects.  

Such a study requires a large enough sample that statistical biases
can average out (e.g. the geometry, inclination, effective radius,
distance, and non homogeneity of the central regions). Furthermore
L$_{1.65\mu m}$ is not indicative of L$_{Bol}$ so that only comparisons 
of M/L derived in the same way are meaningful, and absolute values 
can be misleading. To the best of our knowledge no sample of normal galaxies
has been observed in the near IR using a 1 arcsec {\it slit}.
Previous observations
(e.g. Oliva et al. 1999 and references therein) employed a 3 to 5
arcsec {\it aperture} preventing a direct comparison unless the light
distribution is assumed to be homogeneous, to allow convertion of slit
fluxes to aperture fluxes.   

Thus mass-to-light ratios can be estimated for only two galaxies from 
our sample, NGC~2992 and MCG~-06-30-15, being the only two to have
apparently homogeneous populations surrounding their nucleus (see
Sect. 3). Reddening corrected fluxes within 1.5'' and 4.5'' equivalent
apertures have been derived from the 1.5''x1'' (nucleus) and 4.5''x1'' 
("off-nucleus"+"nucleus") observations. In the case of MCG~-06-30-15, a
50\% stellar contribution was assumed within the nucleus (see
Sect. 3.5). Within the inner regions, quite similar values of 
M/L, ie. 0.074 and 0.065, are found for NGC~2992
and MCG~-06-30-15 respectively. Whereas within
the larger aperture the ratios are M/L=0.047 and 0.12 for NGC~2992
and MCG~-06-30-15
respectively. Since the typical scatter in M to 1.65$\mu$ m luminosity for
galaxies is a factor of 3 to 4 (see Devereux et al. 1987 and Oliva et
al., 1999) no significant difference in age of the stellar population can
be deduced. We also note that, albeit at the same distance, the two
galaxies have very different effective radii (Re = 17.3'' for NGC~2992
and 9'' for MCG~-06-30-15) implying that different amounts of bulge starlight
are enclosed in the fixed apertures. This should be compared with what
can be inferred from direct comparison of the detailed spectra.

The nuclear spectra, normalised at 1.58$\mu$m, of the three Seyfert~2 are
shown in Figure~8. Strong similarities between NGC~3185 and NGC~6221
are clear, whereas the line profiles of NGC~2992 are obviously different.  In
Figure~9 it can be seen that the spectrum of NGC~2992 more closely
ressembles the Seyfert~1 nuclear spectra, and in particular that of
MCG-06-30-15. This is consistent with the fact that NGC~2992 displays
other properties associated with a Seyfert~1.9 rather than a
prototypical type 2. It is noticeable in Figure~9 that NGC~3783 and
MCG-06-30-15 have different stellar populations even though these features
are diluted; e.g. note the features around 1.578$\mu$m (MgI, FeI, CO, OH) and
1.59$\mu$m (SiI, FeI, OH).

\begin{figure*}
\includegraphics[width=6.5cm,bb= 28 28 320 810,angle=-90]{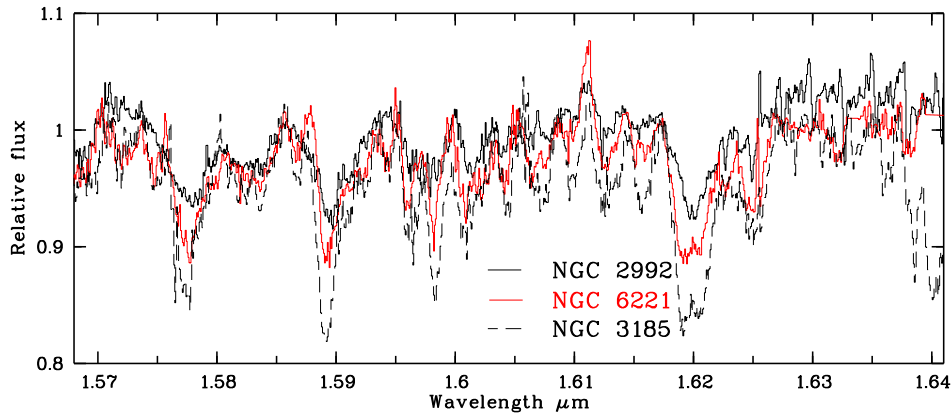}
\caption{Comparison of the nuclear spectra of the Seyfert~2 galaxies.}
\end{figure*}

\begin{figure*} 
\includegraphics[width=6.5cm,bb= 28 28 320 810,angle=-90]{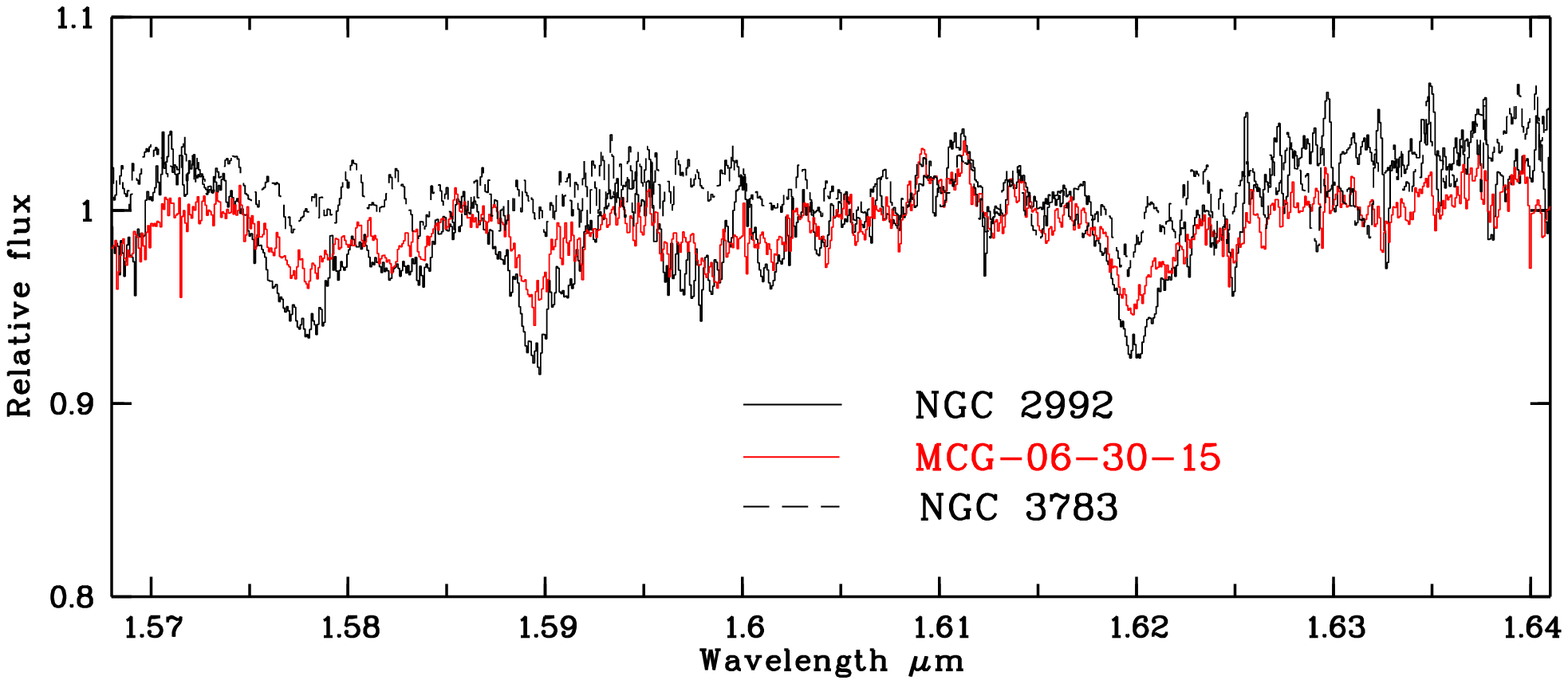} 
\caption{Comparison of the nuclear spectra of
both Seyfert~1 galaxies with that of NGC~2992.}  
\end{figure*}

We can also compare circumnuclear spectra extracted from the same
physical distance (in pc) from the nucleus. In the Seyfert~2 galaxies,
NGC~3185 and NGC~6221, a region at about 120 pc from the centre has
been extracted: both spectra are very similar, suggesting that their
stellar populations are alike, although the morphological type of
their host galaxies are somewhat different (SBa and SBbc,
respectively). In MCG-06-30-15 (an E galaxy) and NGC~2992 (a Sa
galaxy), regions located at 200-220 pc also show similarities in their 
spectra. While the two regions 270 pc away from the nucleus observed
in NGC~3783 (a Seyfert~1) and NGC~6221 (a Seyfert~2) are significantly
different, reflecting different stellar populations due either to the
different morphological type of the host galaxy and/or to the influence of
their different AGN type. We note that in the visible range, Boisson et al.
(2000) and Joly et al. (2001) found the stellar population of the nucleus to be
related to the level of AGN activity, and that the circumnuclear populations
within the central 500 pc were similar to, but younger than, those
typical of a normal galaxy.

\appendix

\section{Stars.}

A set of stars has been observed under the same conditions as for the
galaxies. These stars extend the stellar library
published in Dallier et al. (1996).
Figures~A1 and A2 show the individual flux calibrated spectra. 

\begin{figure}
\resizebox{\hsize}{!}{\includegraphics{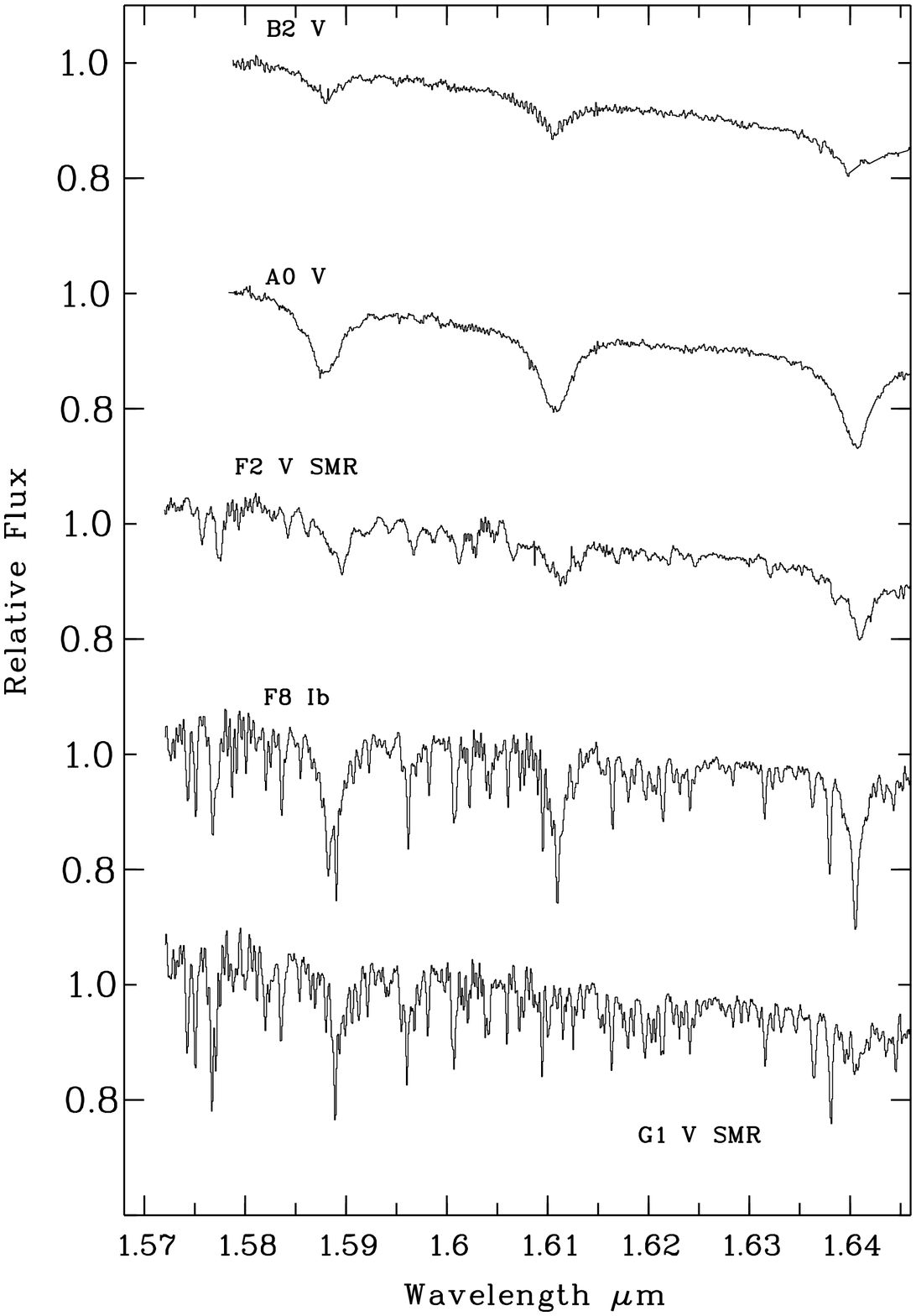}}
\caption{Flux calibrated spectra normalized to 1.,
in the range 1.5852-1.5865$\mu$m, shifted for clarity.  }
\end{figure}

\begin{figure}
\sidecaption
\resizebox{\hsize}{!}{\includegraphics{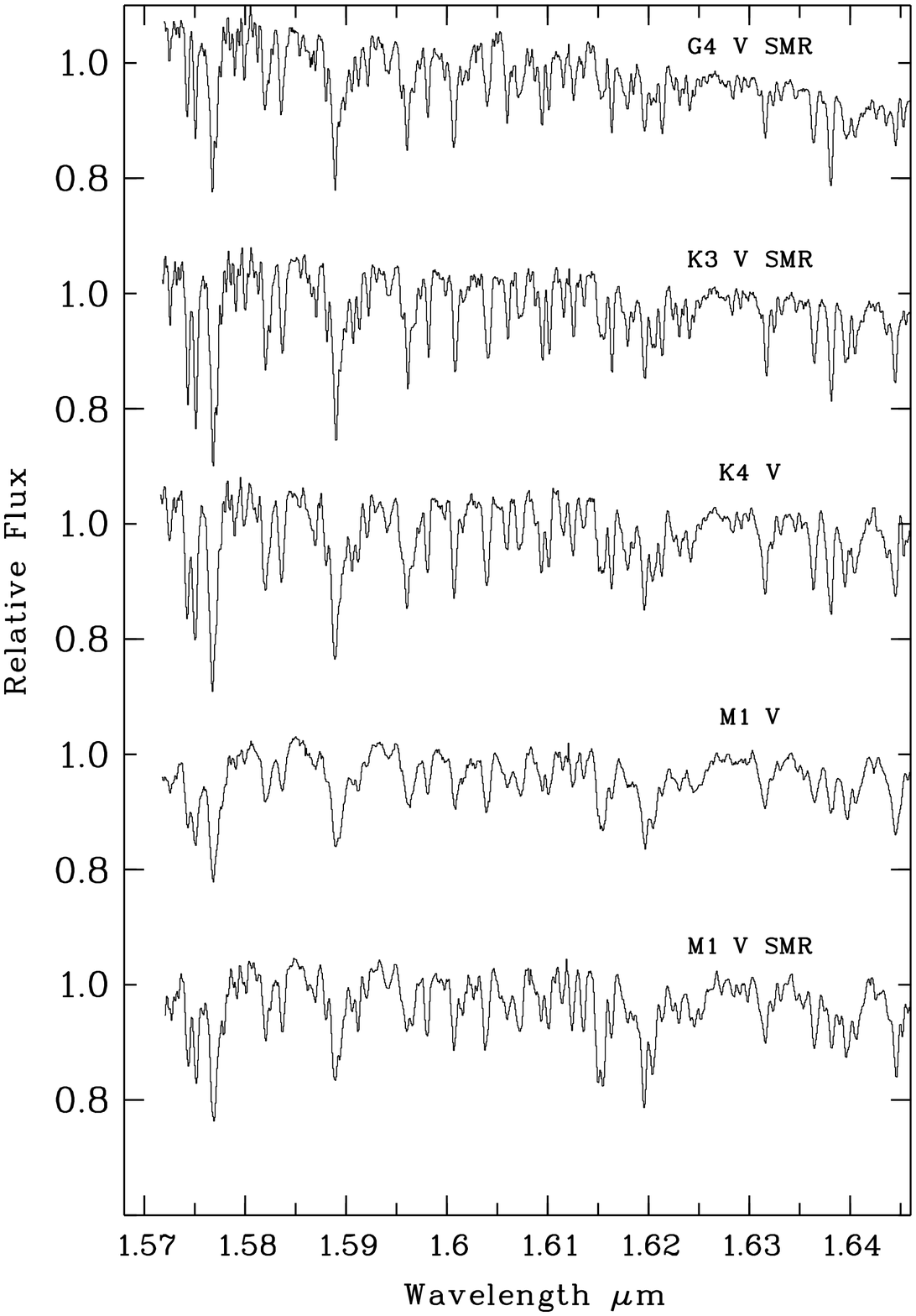}}
\caption{Flux calibrated spectra normalized to 1.,
in the range 1.5852-1.5865$\mu$m, shifted for clarity.  }
\end{figure}

\end{document}